\documentstyle[eqsecnum,aps,epsf
]{revtex}
\newcommand{\DS}{\displaystyle}

\newcommand{\meanrp}[1]{\left\langle\,#1\,\right\rangle^{{\bf p}_0,{\bf x}_0}_{\bf \Omega}}
\newcommand{\meanzero}[1]{\left\langle\,#1\,\right\rangle^{\bf 0}_{\bf \Omega}}

\newcommand{\osa}{\Omega_{\perp 1}}
\newcommand{\osb}{\Omega_{\perp 2}}
\newcommand{\os}{\Omega_{\perp}}
\newcommand{\op}{\Omega_{\parallel}}

\begin{document}
\title{
Quantum Statistics of Hydrogen in Strong Magnetic Fields}
\author{M. Bachmann\thanks{Supported
by the Studienstiftung des deutschen Volkes.}
\thanks{email: mbach@physik.fu-berlin.de}, 
H. Kleinert\thanks{email: kleinert@physik.fu-berlin.de}, and 
A. Pelster\thanks{email: pelster@physik.fu-berlin.de}}
\address{Institut f\"ur Theoretische Physik, Freie Universit\"at Berlin,
Arnimallee 14, 14195 Berlin}
\date{\today}
\maketitle
\begin{abstract}
By an extension of
the Feynman-Kleinert variational approach, we calculate the temperature-dependent
effective classical potential governing the quantum statistical properties of 
a hydrogen atom in a uniform magnetic field.
In the zero-temperature limit, we obtain ground state energies which are
accurate for all magnetic field strengths from weak to strong fields. 
\end{abstract}
\section{Introduction}
The recent discovery of magnetars has renewed interest in the behavior  
of charged particle systems in the presence of extremely strong 
external magnetic fields~\cite{kou}. 
In this new type of neutron stars, electrons and protons 
from decaying neutrons produce magnetic fields $B$ reaching up to $10^{15}\,{\rm G}$,
much larger than those in
neutron stars and white dwarfs, where $B$ is of order $10^{10}-10^{12}\,{\rm G}$
and $10^6-10^8\,{\rm G}$, respectively. 

Analytic treatments of the strong-field properties of an atomic system 
are difficult, even in the zero-temperature limit. 
The reason is the logarithmic asymptotic behavior of the ground state 
energy~\cite{landau,zinn-justin}.
In the weak-field limit, perturbative approaches
~\cite{cizek1,cizek2} yield well-known series expansions 
in powers of $B^2$. These are useful, however, only for $B\ll B_0$, where $B_0$ is the 
atomic magnetic field strength 
$B_0=e^3M^2/\hbar^3\approx 2.35\times 10^5\,{\rm T}=2.35\times10^9\,{\rm G}$.

So far, the most reliable values for strong uniform fields 
were obtained by numerical calculations~\cite{wunner}.
An analytic mapping procedure was introduced in Ref.~\cite{zinn-justin}
to interpolate between the weak- and strong-field behavior, and a 
variational approximation was given in Ref.~\cite{feranshuk}, both with quite good results.

In this note, we use an extension of the Feynman-Kleinert
variational approach~\cite{fk} to find a {\it single analytic 
approximation} to the effective classical potential of the system 
for {\it all} temperatures and magnetic field strengths.
From this, the quantum statistical partition function can be
obtained by a simple 
configuration space integral over a classical-looking Boltzmann-factor.
In the zero-temperature limit, the effective classical potential
is the ground state energy of the system.
\section{Effective Classical Potential}
The Hamiltonian of the electron in a hydrogen atom in the presence 
of a uniform external magnetic field pointing along the positive $z$-axis is
\begin{equation}
\label{ha00}
H({\bf p},{\bf x})=\frac{1}{2M}{\bf p}^2-\frac{1}{2}\omega_cl_z({\bf p},{\bf x})+\frac{1}{8}
\omega_c^2{\bf x}^2-
\frac{e^2}{|{\bf x}|}.
\end{equation}
Here we have used the symmetric gauge ${\bf A}({\bf x})=(B/2)(-y,x,0)$, and denoted
the $z$-component of the orbital angular momentum by 
$l_z({\bf p},{\bf x})=({\bf x}\times {\bf p})_z$. 
The quantum statistical partition function can always be expressed as a classical-looking 
configuration space integral~\cite{fk}
\begin{equation}
\label{ha01}
Z=\int \frac{d^3x_0}{\lambda_{\rm th}^{3}}\,\exp\left[-\beta V_{\rm eff}({\bf x}_0)\right],
\end{equation}
where $\lambda_{\rm th}=\sqrt{2\pi\hbar^2\beta/M}$ is the thermal wavelength, 
$\beta=1/k_BT$ is the inverse temperature, and $V_{\rm eff}({\bf x}_0)$
is the effective classical potential $V_{\rm eff}({\bf x}_0)$. Generalizing
the development in Ref.~\cite{fk}, this is defined by
the phase space path integral
\begin{equation}
\label{ha02}
\exp\left[-\beta V_{\rm eff}({\bf x}_0)\right]\equiv\lambda_{\rm th}^3\int d^3p_0
\oint{\cal D}^3x{\cal D}^3p\,\delta({\bf x}_0-\overline{{\bf x}(\tau)})
\delta({\bf p}_0-\overline{{\bf p}(\tau)})\,e^{-{\cal A}[{\bf p},{\bf x}]/\hbar},
\end{equation}
where ${\cal A}[{\bf p},{\bf x}]$ is the Euclidean action 
\begin{equation}
\label{ha03}
{\cal A}[{\bf p},{\bf x}]=\int_0^{\hbar\beta} d\tau\,[-i{\bf p}(\tau)\dot{\bf x}(\tau)+
H({\bf p}(\tau),{\bf x}(\tau))],
\end{equation}
and $\overline{{\bf x}(\tau)}=\int_0^{\hbar\beta}
d\tau\,{\bf x}(\tau)/\hbar\beta$ and 
$\overline{{\bf p}(\tau)}=\int_0^{\hbar\beta}
d\tau\,{\bf p}(\tau)/\hbar\beta$ are the temporal averages of position and momentum.
The special treatment of ${\bf x}_0$ and ${\bf p}_0$ is necessary, 
since the classical harmonic fluctuation widths $\langle{\bf x}^2 \rangle^{\rm cl}$ and
$\langle{\bf p}^2 \rangle^{\rm cl}$ are proportional to the temperature $T$ (Dulong-Petit law).
Thus they diverge for $T\to\infty$ and their fluctuations cannot be treated pertubatively.
In contrast, the fluctuation widths $\langle({\bf x}-{\bf x}_0)^2\rangle$, 
$\langle({\bf p}-{\bf p}_0)^2\rangle$
around ${\bf x}_0$ and ${\bf p}_0$ go to zero for large $T$ and are limited
down to $T=0$, thus allowing for a treatment by variational
perturbation theory~\cite{PI}. For this
we rewrite the action (\ref{ha03}) as
\begin{equation}
\label{ha05}
{\cal A}[{\bf p},{\bf x}]=
{\cal A}^{{\bf p}_0,{\bf x}_0}_{\bf \Omega}[{\bf p},{\bf x}]
+{\cal A}_{\rm int}[{\bf p},{\bf x}],
\end{equation}
with a harmonic trial action
\begin{eqnarray}
\label{ha06}
{\cal A}^{{\bf p}_0,{\bf x}_0}_{\bf \Omega}[{\bf p},{\bf x}]=
\int_0^{\hbar\beta}d\tau\, \Big\{&&-i[{\bf p}(\tau)-{\bf p}_0]\cdot\dot{\bf x}(\tau)+
\frac{1}{2M}[{\bf p}(\tau)-{\bf p}_0]^2+\frac{1}{2}\osa
l_z({\bf p}(\tau)-{\bf p}_0,{\bf x}(\tau)-{\bf x}_0)\nonumber\\
&&+\frac{1}{8}M\osb^2
\left[{\bf x}^\perp(\tau)-{\bf x}_0^\perp\right]^2+\frac{1}{2}M\op^2[z(\tau)-z_0]^2 \Big\},
\end{eqnarray}
in which ${\bf x}^\perp=(x,y)$ denotes the transverse part of ${\bf x}$. The frequencies 
${\bf \Omega}=(\osa,\osb,\op)$ are arbitrary for the moment. Inserting the decomposition
(\ref{ha05}) into (\ref{ha02}), we expand the exponential of the interaction, 
$\exp\left\{-{\cal A}_{\rm int}[{\bf p},{\bf x}]/\hbar\right\}$, yielding a series of 
expectation values of powers of the interaction
\begin{equation}
\label{vpt07}
\meanrp{{\cal A}_{\rm int}^n[{\bf p},{\bf x}]}=
\frac{(2\pi\hbar)^3}{Z^{{\bf p}_0,{\bf x}_0}_{\bf \Omega}}\oint{\cal D}^3x {\cal D}^3p
\,{\cal A}_{\rm int}^n[{\bf p},{\bf x}]\,\delta({\bf x}_0-\overline{{\bf x}(\tau)})\delta({\bf p}_0-\overline{{\bf p}(\tau)})
\exp\left\{-\frac{1}{\hbar}{\cal A}^{{\bf p}_0,{\bf x}_0}_{\bf \Omega}[{\bf p},{\bf x}]
\right\}.
\end{equation}
The path integral over the Boltzmann-factor involving the harmonic action (\ref{ha06}) 
is exactly solvable and yields the restricted partition function
\begin{equation}
\label{ha08}
Z^{{\bf p}_0,{\bf x}_0}_{\bf \Omega}=
\frac{\hbar\beta\Omega_+/2}{\sinh{\hbar\beta\Omega_+/2}}\,
\frac{\hbar\beta\Omega_-/2}{\sinh{\hbar\beta\Omega_-/2}}\,
\frac{\hbar\beta\op/2}{\sinh{\hbar\beta\op/2}},
\end{equation}
where $\Omega_\pm\equiv|\osa\pm\osb|/2$.
Rewriting the perturbation series as a cumulant expansion, evaluating the expectation 
values, and integrating out the momenta on the right-hand side of Eq.~(\ref{ha02})
leads to a series representation for the effective classical potential $V_{\rm eff}({\bf x}_0)$.
Since it is impossible to sum up the series, the perturbation expansion must
be truncated, leading to an $N$th-order approximation $W^{(N)}_{\bf \Omega}({\bf x}_0)$
for the effective classical
potential. Since the parameters ${\bf \Omega}$ 
are arbitrary, $W^{(N)}_{\bf \Omega}({\bf x}_0)$ should depend {\it minimally} on 
${\bf \Omega}$. This 
determines the optimal values ${\bf \Omega}^{(N)}=(\osa^{(N)}({\bf x}_0),
\osb^{(N)}({\bf x}_0), \op^{(N)}({\bf x}_0))$ of $N$th order.
Reinserting these into $W^{(N)}_{\bf \Omega}({\bf x}_0)$ yields the optimal 
approximation $W^{(N)}({\bf x}_0)\equiv W^{(N)}_{{\bf \Omega}^{(N)}}({\bf x}_0)$.

\begin{figure}
\centerline{
\epsfxsize=12cm \epsfbox{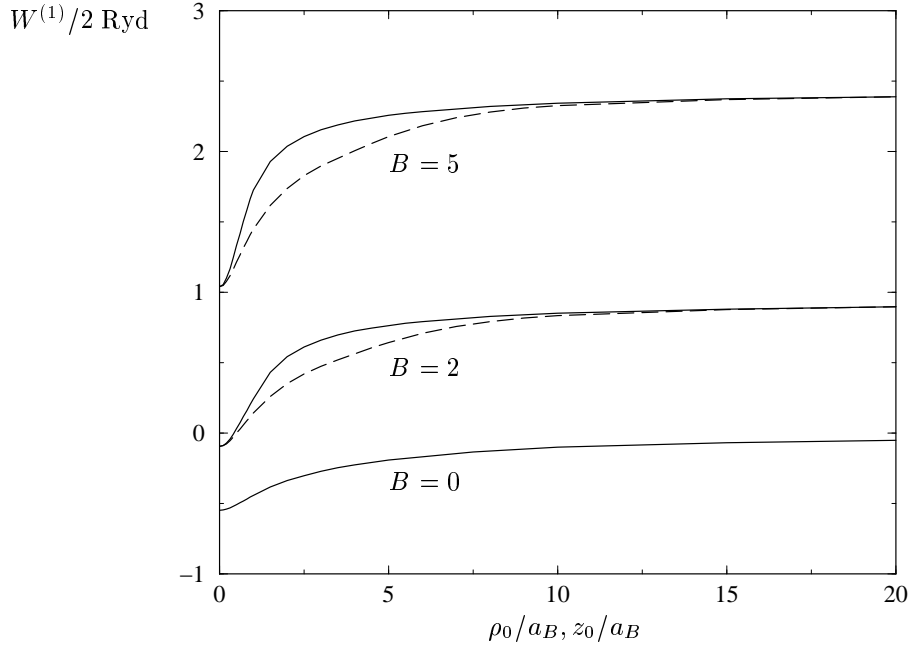}
}
\caption{\label{wofq}
Effective classical potential plotted along along two directions: once as a 
function of the coordinate
$\rho_0=\sqrt{x_0^2+y_0^2}$ perpendicular to the field lines at $z_0=0$ (solid curves),
and once parallel to the magnetic field as a function of $z_0$ at $\rho_0=0$ (dashed curves).
The inverse
temperature is fixed at $\beta=100$, and the strengths of the magnetic field $B$ are varied
(all in natural units).
}
\end{figure}

The first-order approximation to the effective classical potential is
\begin{equation}
\label{ha10}
W_{\bf \Omega}^{(1)}({\bf x}_0)=-\frac{1}{\beta}{\rm ln}\,
Z^{{\bf p}_0,{\bf x}_0}_{\bf \Omega}
+(\omega_c-\osa)\,b^2_\perp({\bf x}_0)-\frac{1}{4}\left(\osb^2-\omega_c^2 \right)
\,a^2_\perp({\bf x}_0)-\frac{1}{2}M\op^2a^2_\parallel({\bf x}_0)-
\meanrp{\frac{e^2}{|{\bf x}|}},
\end{equation}
where the quantities $a^2_\perp({\bf x}_0)$, $b^2_\perp({\bf x}_0)$, and 
$a^2_\parallel({\bf x}_0)$ are the transverse and longitudinal fluctuation widths
\begin{equation}
\label{ha11}
a^2_\perp({\bf x}_0)=\meanrp{x^2(\tau)},
\quad a^2_\parallel({\bf x}_0)=\meanrp{z^2(\tau)},
\quad b^2_\perp({\bf x}_0)=\meanrp{x(\tau)p_y(\tau)}.
\end{equation}
The expectation value of the Coulomb potential on the right-hand side of Eq.~(\ref{ha10})
has the integral representation
\begin{eqnarray}
\label{ha12}
-\meanrp{\frac{e^2}{|{\bf x}|}}&=&-e^2
\sqrt{\frac{2}{\pi}\,a^2_\parallel({\bf x}_0)}\int\limits_0^1\frac{d\xi}
{a^2_\parallel({\bf x}_0)+\xi^2[a^2_\perp({\bf x}_0)-a^2_\parallel({\bf x}_0)]}\nonumber\\
&&\times\exp\left\{-\frac{\xi^2}{2}\left(\frac{x_0^2+y_0^2}{a^2_\parallel({\bf x}_0)+
\xi^2[a^2_\perp({\bf x}_0)-a^2_\parallel({\bf x}_0)]}+\frac{z_0^2}{a^2_\parallel({\bf x}_0)}
\right) \right\}.
\end{eqnarray}
The variational energy (\ref{ha10}) is minimized at each ${\bf x}_0$, and the resulting
$W^{(N)}({\bf x}_0)$ is displayed for a low temperature and different magnetic fields
in Fig.~\ref{wofq}.

From now on we set $\hbar=e^2=k_B=c=M=1$. 
Thus, energies are measured in units of
$\epsilon_0=Me^4/\hbar^2\equiv 2\,{\rm Ryd}\approx 27.21\,{\rm eV}$,
temperatures in
$\epsilon_0/k_B\approx 3.16\times 10^5\,{\rm K}$, distances in Bohr radii
$a_B=\hbar^2/Me^2\approx
0.53\times 10^{-8}\,{\rm cm}$, and magnetic field strengths in
$B_0=e^3M^2/\hbar^3\approx 2.35\times 10^5\,{\rm T}=2.35\times10^9\,{\rm G}$.
The curves $W^{(1)}({\bf x}_0)$ are plotted to show their
anisotropy with respect to the magnetic field direction.
The anisotropy grows when 
lowering the temperature and increasing the field strength.
Far away from the proton at the origin, the potential becomes isotropic, 
due to the decreasing influence of the Coulomb interaction. 
Analytically, this is seen by going to the limits
$\rho_0\to\infty$ or $z_0\to\infty$, where the expectation value of the Coulomb potential
(\ref{ha12}) tends to zero, leaving an effective classical potential
\begin{equation}
\label{ha13}
W_{\bf \Omega}^{(1)}({\bf x}_0)\longrightarrow -\frac{1}{\beta}{\rm ln}\,
Z^{{\bf p}_0,{\bf x}_0}_{\bf \Omega}
+(\omega_c-\osa)\,b^2_\perp-\frac{1}{4}\left(\osb^2-\omega_c^2 \right)
\,a^2_\perp-\frac{1}{2}M\op^2a^2_\parallel.
\end{equation}
This is ${\bf x}_0$-independent,
and optimization yields the constants $\osa^{(1)}=\osb^{(1)}=\omega_c$
and $\op^{(1)}=0$, with the asymptotic energy
\begin{equation}
\label{ha14}
W^{(1)}({\bf x}_0)\longrightarrow -\frac{1}{\beta}{\rm ln}\,
\frac{\hbar\beta\omega_c/2}{\sinh{\hbar\beta\omega_c/2}}.
\end{equation}
The $B=0$~-curves are, of course, identical with those obtained from variational
perturbation theory for the hydrogen atom~\cite{hatom}.

For large temperatures,
the anisotropy decreases since the violent thermal
fluctuations have a smaller preference of the $z$-direction. 
\section{Zero-Temperature Limit}
\label{vptlow}
At zero-temperature, the first-order effective classical 
potential (\ref{ha10}) at the origin yields an approximation for the  
ground state energy of the hydrogen atom in a uniform magnetic field: 
$E_{\bf \Omega}^{(1)}=\lim_{\beta\to\infty}W_{\bf \Omega}^{(1)}({\bf 0})$: 
\begin{equation}
\label{gr01}
E^{(1)}_{\bf \Omega}(B)=\frac{1}{4\osb}\left(\osb^2+\omega_c^2\right)+\frac{\op}{4}-
\meanzero{\frac{1}{|{\bf x}|}},
\end{equation}
with the expectation value for the Coulomb potential
\begin{equation}
\label{gr02}
\meanzero{\frac{1}{|{\bf x}|}}=\frac{2}{\sqrt{\pi}}\times\left\{
\begin{array}{cl}
\DS\sqrt{\frac{\op\osb}{\op-\osb}}\,{\rm arctan}\,\sqrt{\frac{2\op}{\osb}-1}, & 
\quad 2\op > \osb,\\[2mm]
\DS\sqrt{\op}, & \quad 2\op = \osb,\\[2mm]
\DS\frac{1}{2i}\sqrt{\frac{\op\osb}{\op-\osb}}\,{\rm ln}\,
\frac{1+i\sqrt{2\op/\osb-1}}{1-i\sqrt{2\op/\osb-1}}, & \quad 2\op < \osb.
\end{array}\right.
\end{equation}
Equations (\ref{gr01}) and (\ref{gr02}) are independent of the frequency parameter
$\osa$, such that optimization of the ground state energy
(\ref{gr01}) is ensured by minimization.
Reinserting the extremal $\osb^{(1)}$ and $\op^{(1)}$ into Eq.~(\ref{gr01})
yields the first-order approximation to the ground state energy $E^{(1)}(B)$. In the absence
of
the Coulomb interaction the optimization with respect to $\osb$ yields $\osb^{(1)}=\omega_c$,
rendering the ground state energy $E^{(1)}(B)=\omega_c/2$, which
is the zeroth Landau level in this special case. 
The trial frequency $\op$ must be set equal to zero to preserve translational symmetry
along the $z$-axis. 

In the opposite limit of a vanishing magnetic field, 
Eq.~(\ref{gr01}) coincides with the first-order
variational result for the ground state energy of the hydrogen atom,
whose optimization gave
$E^{(1)}(B=0)=-4/3\pi\approx -0.4244\, {\rm [2\,Ryd]}$ obtained in Refs.~\cite{fk,PI}.
In Ref.~\cite{hatom}, the $B=0$~-system was treated up to third order leading to the 
much more accurate
result $E^{(1)}(B=0)\approx -0.490\, {\rm [2\,Ryd]}$, very near the exact value 
$E^{\rm ex}(B=0)=-0.5 \, {\rm [2\,Ryd]}$.

\begin{figure}
\centerline{\epsfxsize=12cm \epsfbox{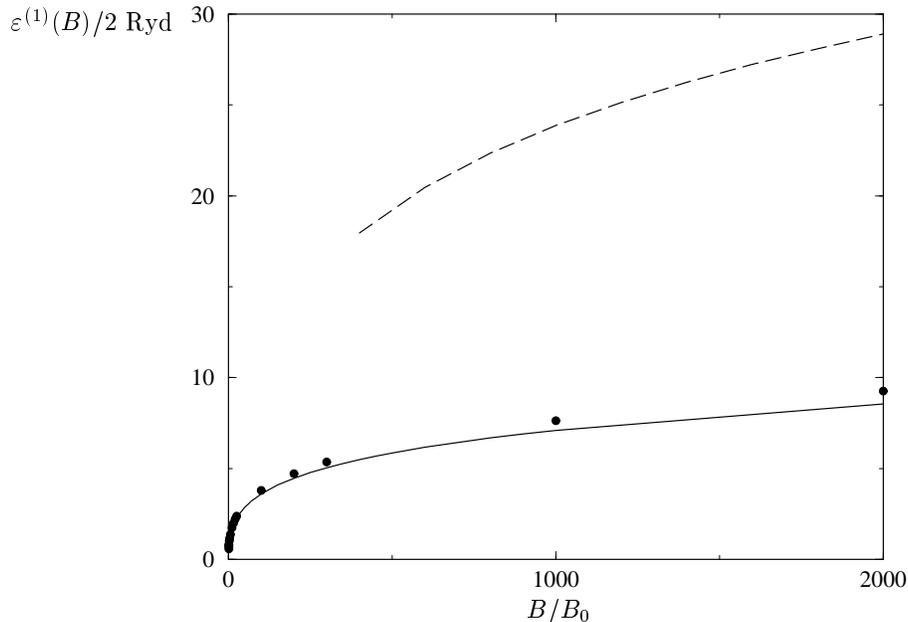}}
\caption[]{\label{bofb} First-order variational result for the binding energy (\ref{int00})
as a function of the
strength of the magnetic field. The dots indicate the values of Ref.~\cite{zinn-justin}.
The dashed curve shows the simple estimate $0.5\, {\rm ln}^2B$ 
of Landau-Lifschitz~\cite{landau}.}
\end{figure}

Let us investigate the asymptotics in the strong-field limit $B\to\infty$. The $B$-dependence
of the binding energy
\begin{equation}
\label{int00}
\varepsilon(B)=\frac{B}{2}-E
\end{equation}
is plotted in Fig.~\ref{bofb}, where it is compared with the results of
Ref.~\cite{zinn-justin}, with satisfactory agreement.
Our results are of similar accuracy as those of 
other first-order calculations, for example those from the operator
optimization method in first order of Ref.~\cite{feranshuk}. The advantage
of variational perturbation theory is that it yields good results 
for all magnetic field strengths. From our experience with the fast    
convergence of the method~\cite[Chaps. 5,9]{PI}, higher orders
of variational perturbation theory will push the approximations
rapidly towards the exact value.
\subsection{Weak-Field Behavior}
The calculations of the binding energy for weak magnetic fields show that the ratio
$\eta\equiv 2\op/\osb$ is always smaller than one if $B\neq0$. Setting
$\Omega\equiv\osb$, we rewrite the binding energy as a function of $\Omega$ and $\eta$:
\begin{equation}
\label{we00}
\varepsilon^{(1)}_{\eta,\Omega}(B)\approx
\frac{B}{2}-\frac{\Omega}{4}\left(1+\frac{\eta}{2}\right)-\frac{B^2}{4\Omega}
-\sqrt{\frac{\eta\Omega}{2\pi}}\frac{1}{\sqrt{1-\eta}}\,{\rm ln}\,\frac{1-\sqrt{1-\eta}}
{1+\sqrt{1-\eta}}.
\end{equation} 
This is minimized with
respect to the new variational parameters $\eta$ and $\Omega$
by expanding $\eta(B)$ and $\Omega(B)$ in powers of $B^2$ with unknown coefficients, and 
inserting these expansions
into extremality equations. 
The expansion coefficients are then determined order by order.
The optimal expansions are inserted into (\ref{we00}), yielding
the optimized binding energy 
$\varepsilon^{(1)}(B)$ as a power series
\begin{equation}
\label{we06}
\varepsilon^{(1)}(B)=\frac{B}{2}-\sum\limits_{n=0}^\infty\,\varepsilon_n B^{2n}.
\end{equation}
The coefficients $\varepsilon_n$ are listed in Table~\ref{tab1} and compared with the
exact ones of Ref.~\cite{cizek1}. Of course,
the higher-order coefficients of this first-order variational approximation
become rapidly inaccurate,
but the results can be improved, if desired, by going to higher orders in 
variational perturbation theory as in Ref.~\cite[Chaps. 5,9]{PI}.
\subsection{Strong-Field Behavior}
In the discussion of the pure magnetic field we have mentioned
that the variational calculation for the ground state energy, which is associated
with the zeroth Landau level, yields a frequency $\osb\propto B$, while $\op=0$. We therefore
use the assumptions
$\os\equiv \osb \gg 2\op$ and $\op \ll B$
for an analytic study of the strong-field behavior of the ground state energy (\ref{gr01}).
We expand the last expression of the expectation value (\ref{gr02}) 
in terms of $2\op/\os$, and reinsert this expansion into (\ref{gr01}).
Then we omit all terms proportional to $C/\os$, where $C$ stands for any expression
with a value much smaller than the field strength $B$. We thus
obtain the strong-field approximation for the first-order binding energy (\ref{int00})
\begin{equation}
\label{st01}
\varepsilon_{\os,\op}^{(1)}=\frac{B}{2}-\left(\frac{\os}{4}+\frac{B^2}{4\os}+\frac{\op}{4}
+\sqrt{\frac{\op}{\pi}}\,{\rm ln}\,\frac{\op}{2\os}\right).
\end{equation}
Determining $\os$, $\op$ by minimization, we obtain
\begin{equation}
\label{st02}
\os\approx B,\qquad \sqrt{\op^{(3)}}= \frac{2}{\sqrt{\pi}}
\left( {\rm ln}\,B -2{\rm ln}{\rm ln}\,B+
\frac{2a}{{\rm ln}\,B}+\frac{a^2}{{\rm ln}^2B}+b \right) +{\cal O}({\rm ln}^{-3} B)
\end{equation} 
with abbreviations
$a=2-{\rm ln}\,2 \approx 1.307$ and $b= {\rm ln}(\pi/2)-2\approx -1.548$.
Thus the optimized binding energy can be written up to the
order ${\rm ln}^{-2}B$:
\begin{eqnarray}
\label{st15}
\varepsilon^{(1)}(B)&=&\frac{1}{\pi}\left\{{\rm ln}^2B -4\,{\rm ln}\,B\; {\rm ln}{\rm ln}\,B
+4\,{\rm ln}^2{\rm ln}\,B -4b\,{\rm ln}{\rm ln}\,B
+2(b+2)\,{\rm ln}\,B
+b^2
-\frac{1}{{\rm ln}\,B}\left[8\,{\rm ln}^2{\rm ln}\,B-8b\,{\rm ln}{\rm ln}\,B+2b^2
\right]\right\}\nonumber\\
&&+{\cal O}({\rm ln}^{-2}B).
\end{eqnarray}
Note that the prefactor $1/\pi$ of the leading ${\rm ln}^2B$-term
differs from a value $1/2$ obtained by Landau and Lifschitz~\cite{landau}.
Our value is a consequence of the harmonic trial system.
The calculation of higher orders
in variational perturbation theory should drive our value
towards $1/2$.

The convergence of the expansion (\ref{st15}) is quite slow. 
At a magnetic field strength $B=10^5 B_0$, which corresponds to $2.35\times 10^{10}\,{\rm T}=
2.35\times10^{14}\,{\rm G}$,
the contribution from the first six terms is $22.87\,[2\,{\rm Ryd}]$.
The next three terms suppressed
by a factor ${\rm ln}^{-1} B$ contribute $-2.29\,[2\,{\rm Ryd}]$,
while an estimate for the ${\rm ln}^{-2} B$-terms
yields nearly $-0.3\,[2\,{\rm Ryd}]$. Thus we find 
$\varepsilon^{(1)}(10^5)=20.58\pm 0.3\,[2\,{\rm Ryd}]$.
This is in very good agreement with the value $20.60\,[2\,{\rm Ryd}]$
obtained from the full treatment described in Sec.~\ref{vptlow}.

Table~\ref{asymp} lists the values of the first six terms of Eq.~(\ref{st15}).
This shows in particular the significance of the second term
in (\ref{st15}), which is of the same order of the leading first term,
but with an opposite sign. In Fig.~\ref{bofb},
we have plotted the expression
$\varepsilon_L(B)=(1/2)\,{\rm ln}^2B$
of Landau and Lifschitz~\cite{landau} to illustrate that it gives far too large
binding energies even at very large magnetic fields, e.g. at $2000 B_0\propto 10^{12}\,{\rm G}$.
Obviously, the nonleading terms in Eq.~(\ref{st15})
give important contributions to the asymptotic behavior even at such large magnetic fields.
As an peculiar property of the asymptotic behavior, the absolute value of
the difference between the Landau-expression $\varepsilon_L(B)$ 
and our approximation (\ref{st15})
diverges with increasing magnetic field strengths $B$. Only the relative difference
decreases.
\section{Summary}
We have calculated the effective classical potential for the hydrogen atom 
in constant magnetic field, which governs the statistical mechanics of the system
at all temperatures. 
At zero temperature, we find a rather accurate ground state energy
which interpolates very well between weak and strong fields. 
\begin{table}
\caption[]{\label{tab1} Perturbation coefficients up to order $B^6$ for the weak-field
expansions of the variational parameters and the binding energy in comparison to the
exact ones of Ref.~\cite{cizek1}.}
\begin{tabular}{c|cccc}
$n$ & 0 & 1 & 2 & 3\\ \hline\hline
\rule[-8pt]{0pt}{23pt} $\eta_n$ & $1.0$ & $\DS-\frac{405\pi^2}{7168}\approx -0.5576$ &
$\DS\frac{16828965\pi^4}{1258815488}\approx 1.3023$ &
$\DS-\frac{3886999332075\pi^6}{884272562962432}\approx -4.2260$\\ \hline
\rule[-8pt]{0pt}{23pt}$\Omega_n$ & $\DS\frac{32}{9\pi}\approx 1.1318$ & $\DS\frac{99\pi}{224}\approx 1.3885$ &
$\DS -\frac{1293975\pi^3}{19668992}\approx -2.03982$ &
$\DS\frac{524431667187\pi^5}{27633517592576}\approx 5.8077$ \\ \hline
\rule[-8pt]{0pt}{23pt}$\varepsilon_n$ & $\DS -\frac{4}{3\pi}\approx -0.4244$ & $\DS\frac{9\pi}{128}\approx 0.2209$ &
$\DS -\frac{8019\pi^3}{1835008}\approx -0.1355$ & $\DS\frac{256449807\pi^5}{322256764928}
\approx 0.2435$\\ \hline
\rule[-8pt]{0pt}{23pt}$\varepsilon_n$~\cite{cizek1} & $-0.5$ & $0.25$ & $\DS -\frac{53}{192}\approx -0.2760$ &
$\DS\frac{5581}{4608}\approx 1.2112$
\end{tabular}
\end{table}
\begin{table}
\caption{\label{asymp} Example for the competing leading six terms in Eq.~(\ref{st15})
at $B=10^5B_0\approx 2.35\times 10^{14}\,{\rm G}$.}
\begin{tabular}{cccccc}
$(1/\pi){\rm ln}^2B$ & $-(4/\pi){\rm ln}\,B\;{\rm ln}{\rm ln}\,B$ &
$(4/\pi)\,{\rm ln}^2{\rm ln}\,B$ &
$-(4b/\pi)\,{\rm ln}{\rm ln}\,B$ & $[2(b+2)/\pi]\,{\rm ln}\,B$ & $b^2/\pi$\\ \hline
\rule[-4pt]{0pt}{15pt}$42.1912$ & $-35.8181$ & $7.6019$ & $4.8173$ & $3.3098$ & $0.7632$
\end{tabular}
\end{table}

\begin{thebibliography}{199}
%
\bibitem{kou}
C.~Kouveliotou et al., Nature {\bf 393}, 235 (1998);
C.~Kouveliotou et al., ApJ {\bf 510}, L115 (1999);
K.~Hurley et al., ApJ {\bf 510}, L111 (1999);
V.M.~Kaspi, D.~Chakrabarty, and J.~Steinberger, ApJ {\bf 525}, L33 (1999);
B.~Zhang and A.K.~Harding, eprint: astrp-ph/0004067 (2000).
%
\bibitem{landau}
L.D.~Landau and E.M.~Lifschitz, {\em Quantenmechanik}, Akademie-Verlag Berlin, Sechste Auflage,
1979.
%
\bibitem{zinn-justin}
J.C.~Le~Guillou and J.~Zinn-Justin, Ann. Phys. (N.Y.) {\bf 147}, 57 (1983).
%
\bibitem{cizek1}
J.E.~Avron, B.G.~Adams, J.~\v{C}\'\i\v{z}ek, M.~Clay, M.L.~Glasser, P.~Otto, J.~Paldus,
and E.~Vrscay, Phys. Rev. Lett. {\bf 43}, 691 (1979).
%
\bibitem{cizek2}
B.G.~Adams, J.E.~Avron, J.~\v{C}\'\i\v{z}ek, P.~Otto, J.~Paldus, R.K.~Moats, and H.J.~Silverstone,
Phys. Rev. A {\bf 21}, 1914 (1980).
%
\bibitem{wunner}
H.~Ruder, G.~Wunner, H.~Herold, and F.~Geyer, {\em Atoms in Strong Magnetic Fields}
(Springer-Verlag, Berlin, 1994).
%
\bibitem{feranshuk}
I.D.~Feranshuk and L.I.~Komarov, J. Phys. A: Math. Gen. {\bf 17}, 3111 (1984).
%
\bibitem{fk}
R.P.~Feynman and H.~Kleinert, Phys. Rev. A {\bf 34}, 5080 (1986); see also
Chapter 5 of the textbook~\cite{PI}. 
%
\bibitem{PI}
H.~Kleinert, 
{\em Path Integrals in Quantum Mechanics, Statistics, and Polymer Physics},
2nd ed. (World Scientific, Singapore, 1995).
%
\bibitem{hatom}
H.~Kleinert, W.~K\"urzinger, and A.~Pelster, J. Phys. A: Math. Gen. {\bf 31}, 8307 (1998),
eprint: quant-ph/9806016.
%
\end{thebibliography}
\end{document}